\documentclass[11pt]{article}
\usepackage{float}
\usepackage{mymacros}
\usepackage{chngcntr}
\usepackage{verbatim}
\usepackage{amsthm}
\usepackage{graphics}
\usepackage{graphicx}
\usepackage{xcolor}
\usepackage{amsmath}
\usepackage{amssymb}
\usepackage{mathrsfs}
\usepackage{bbold}
\usepackage{cases}
\usepackage{epsfig}
\usepackage{epstopdf}
\usepackage{hyperref}

\title{\boldmath Chaos and Complexity for Inverted Harmonic Oscillators}

\author{Le-Chen Qu $^{a\,b}$ \footnote{qulch20@lzu.edu.cn}}
\author{Jing Chen$^{a\,b}$ \footnote{jchen2017@lzu.edu.cn}}
\author{Yu-Xiao Liu$^{a\,b}$
\footnote{liuyx@lzu.edu.cn, corresponding author}}

\affiliation{
$^{a}$Lanzhou Center for Theoretical Physics, Key Laboratory of Theoretical Physics of Gansu Province, School of Physical Science and Technology, Lanzhou University, Lanzhou 730000, China\\
$^{b}$Institute of Theoretical Physics \& Research Center of Gravitation, Lanzhou University, Lanzhou 730000, China
}

\abstract{We investigate the circuit complexity and Loschmidt echo for the (inverted) harmonic oscillators. Focusing on the chaotic behaviors under the perturbation, we analytically derive the Lyapunov exponent and scrambling time of the inverted harmonic oscillators. We show that the circuit complexity and Loschmidt echo exhibit qualitatively similar behaviors, particularly the consistent Lyapunov exponent.}

\begin{document}
\maketitle
\section{Introduction }
The precise definition of quantum chaos is still an open question. Unlike classical chaos, one can only describe different aspects of quantum chaos \cite{{1},{Jahnke:2018off}}. The precise definition of quantum chaos will help the understanding of thermalization, transport in quantum many-body systems and black hole information loss. Therefore, how to accurately define quantum chaos is indeed an important issue \cite{{1},{Jahnke:2018off}}.

The reason for the appearance of classical chaos is that the evolution of the system is very sensitive to the initial conditions due to the highly non-linearity of the equation of motion. And the distance between two adjacent points in the phase space increases as $e^{\lambda_Lt}$, where $\lambda_L$ is the Lyapunov exponent \cite{F. Haake}. However, since Schr\"{o}dinger equation is a linear equation, the evolution of a quantum system is not highly sensitive to the initial state in principle. This forces us to develop new chaotic probes for quantum systems.

Traditionally, the chaos of a quantum system is characterized by random matrix theory \cite{Bohigas:1983er}. Recently inspired by the AdS/CFT duality, out-of-time-order correlator (OTOC) has become a candidate for describing quantum chaos \cite{Maldacena:2016hyu,Larkin1969QuasiclassicalMI,Kitaev:KITP}. Especially in Ref.~\cite{Maldacena:2015waa} Maldacena and collaborators proved the inequality for large $N$ quantum theories at temperature $T$,
\begin{equation}
\label{e20}
\lambda_{L}(T) \leq 2 \pi T,
\end{equation}
where $\lambda_{\mathrm{L}}(T)$ is the Lyapunov exponent in thermal OTOC. When the quantum system is dual to a gravitational system, the inequality is saturated. One of the most famous is Sachdev-Ye-Kitaev model \cite{Sachdev:1992fk}, which is expected to be dual to two-dimensional Jackiw-Teitelboim gravity \cite{Kitaev:2017awl}.

However, a recent work got acquainted with the tense relationship between OTOC and random matrix theory diagnostics \cite{Nosaka:2018iat}. One of the most striking questions is whether it is possible to use quantum information tools to describe quantum chaos \cite{Miyaji:2016fse,Hosur:2015ylk,Roberts:2016hpo,Cotler:2017jue,Magan:2018nmu,Balasubramanian:2019wgd,Yosifov:2019gwt}, such as circuit complexity \cite{Jefferson:2017sdb,Chapman:2017rqy,Jiang:2018nzg,Camargo:2018eof,Ali:2018fcz,Ali:2018aon,Chapman:2018hou,Bhattacharyya:2019kvj,Hackl:2018ptj,Khan:2018rzm,Caceres:2019pgf,Bhattacharyya:2018bbv,Guo:2018kzl,Doroudiani:2019llj,Bai:2021ldj,Guo:2020dsi}. Recent works have shown that circuit complexity can indeed detect quantum chaos \cite{Ali:2019zcj,Bhattacharyya:2020art,Bhattacharyya:2020iic}. In particular, it was found that chaotic behavior can be characterized by the complexity of bidirectional evolution. That is, first evolve a reference state forward with a Hamiltonian $\hat{H}$ and then evolve backward with a slightly different Hamiltonian $\hat{H}+\hat{\delta H}$. Finally, examine the complexity between the resulting state and the chosen reference state.

However, because the expression of complexity is usually very complicated, only the numerical fitting of Lyapunov exponent and scrambling time was obtained in Ref.~\cite{Ali:2019zcj}. In this work, by considering infinitesimal perturbation, we find the analytical expression of the circuit complexity and derive the Lyapunov exponent and scrambling time. As the most natural concept to measure the distance between states, the Loschmidt echo was proposed in Ref.~\cite{Jalabert2001} to describe quantum chaos. Similar to complexity, Loschmidt echo is defined as the inner product under bidirectional evolution. We also find that the Lyapunov exponent can indeed be obtained through the Loschmidt echo, which is consistent with Ref.~\cite{Jalabert2001}.

To obtain the analytical expression of circuit complexity, we investigate the simplest unstable system --- the inverted oscillator, which is described by the Hamiltonian $\hat{H}=\frac{\hat{p}^2}{2m}-\frac{1}{2} m \omega^2 \hat{q}^2$ \cite{Ali:2019zcj}. Classically, the inverted oscillator has an unstable point in phase space at $(x=0, p=0)$; a particle accelerates exponentially away from the fixed point when perturbed. We obtain the complexity and inner product of the inverted harmonic oscillator and compare their difference with that of the harmonic oscillator. We find that the Lyapunov exponent is proportional to the frequency, and the scrambling time will tend to infinity as the amplitude of the perturbation decreases. Using the circuit complexity and Loschmidt echo, we also examine the Lyapunov exponent and scrambling time in a quantum field theory that can be discretized into harmonic oscillators and inverted harmonic oscillators.

The rest of the paper is organized as follows. In section 2, we review some background material and preliminaries. In section 3, we discuss the chaotic behavior of the inverted harmonic oscillator. In section 4, we investigate the chaotic behaviors in quantum field theory via the same tools. Finally, a summary is presented in section 5.

\section{Background Material and Preliminaries}
\subsection{Inverted Harmonic Oscillator}
We start with the Hamiltonian of the inverted harmonic oscillator, namely
\begin{equation}
\label{e1}
\begin{aligned}
\hat{H}=\frac{\hat{p}^2}{2m}+\hat{V},\quad \hat{V}=-\frac{1}{2} m \omega^2 \hat{q}^2,
\end{aligned}
\end{equation}
where $m$ and $\omega$ are the mass and frequency of the inverted harmonic oscillator, respectively. In contrast to the harmonic oscillator, the potential $\hat{V}$ associated with the inverted harmonic oscillator is always negative. Due to the unbounded bottom of the potential, the inverted oscillator does not have a ground state or a complete set of square-integrable energy eigenstates. Strictly speaking, $\hat{V}$ is not a Hermitian operator because it tends to negative infinity at infinity. However, we can usually regard $\hat{V}$ as a local approximation of some Hermitian operators, such as
\begin{equation}
\hat{V'}=-\frac{1}{2} m \omega^2 \hat{q}^2+\lambda\hat{q}^4.
\end{equation}
Obviously, ${V}$ is an approximation of ${V'}$ in the limit $ q \to 0$. Conversely, for the state that is localized at the origin, we can also use $\hat{V}$ instead of $\hat{V'}$ in the Schr\"odinger equation, as long as the system evolution time is short enough. And since $\hat{V}$ is the quadratic form of $\hat{q}$, we can analytically study its many properties. This is also why the inverted harmonic oscillator is widely studied \cite{Barton:1984ey,Betzios:2016yaq,Betzios:2020wcv,Berry,Morita:2019bfr,Morita:2018sen,Bzowski:2018aiq}.

\subsection{Out-of-Time-Order Correlators}
In this section, we review some results of the OTOC in quantum mechanics. Taking the canonical operators $\hat{p}, \hat{q}$, the OTOC is defined as \cite{35,Hashimoto:2017oit}
\begin{equation}
\label{e22}
C_{T}(t)=-\left\langle[\hat{q}(t), \hat{p}(0)]^{2}\right\rangle_\beta\qquad
\text{with} \qquad \hat{q}(t)=e^{i \hat{H} t}\hat{q} e^{-i \hat{H} t}\,,
\end{equation}
where $\langle \cdots\rangle_\beta$ denotes the thermal expectation value, \ie $\langle\mathcal{\hat{O}}\rangle_\beta = \operatorname{tr}\left[e^{-\beta \hat{H}} \mathcal{\hat{O}}\right] /  \operatorname{tr} e^{-\beta \hat{H}}
$ with $\beta=\frac{1}{T}$ as the inverse temperature of the thermal state.
In terms of a set of energy eigenstate basis, we can rewrite the OTOC as \cite{Hashimoto:2017oit}
\begin{equation}
\label{e21}
C_{T}(t)=\frac{\sum_{r} e^{-\beta E_{r}} c_{r}(t)}{\operatorname{tr} e^{-\beta \hat{H}}} , \quad c_{r}(t)=-\left\langle r\left|[\hat{q}(t),\hat{ p}]^{2}\right| r\right\rangle,
\end{equation}
where the energy eigenstates denoted by $|r\rangle$ satisfy $\hat{H}|r\rangle    =E_{r}|r\rangle$. We can apply  the completeness relation $1=\sum_{s}|s\rangle\langle s|$ to obtain the decomposition of the coefficients $c_r(t)$, \ie
\begin{equation}
\label{e14}
c_{r}(t)=\sum_s b_{r s}(t) b_{r s}^{*}(t), \quad b_{r s}(t) =-i\langle r|[\hat{q}(t),\hat{ p}]| s\rangle.
\end{equation}
By defining the matrix elements $q_{rs} =\langle r|\hat{q}| s\rangle$ and $p_{r s} =\langle r|\hat{p}| s\rangle$, it is straightforward to find
\begin{equation}\label{eq:brs}
b_{r s}(t)=-i \sum_{k}\left(e^{i E_{r k} t} q_{r k} p_{k s}-e^{i E_{k s} t} p_{r k} q_{k s}\right),
\end{equation}
with $E_{r s}=E_{r}-E_{s}$. As a result, we would be able to use the elements $p_{rs}, q_{rs}$ and the spectrum $E_r$ to derive the value of the OTOC after choosing a specific Hamiltonian that we are interested in.

First of all, we begin with the normal harmonic oscillator whose Hamiltonian is given by
\begin{equation}\label{eq:HOHamiltonian}
\begin{aligned}
\hat{H}=\frac{\hat{p}^2}{2m}+\frac{1}{2} m \omega^2 \hat{q}^2
\end{aligned}
\end{equation}
with positive potential. One can easily get
\begin{equation}
E_{r}=\left(r+\frac{1}{2}\right) \omega, \quad q_{r s}=\frac{1}{\sqrt{2\omega m}}\left(\sqrt{s} \delta_{r, s-1}+\sqrt{s+1} \delta_{r, s+1}\right).
\end{equation}
By using the commutator relation $[\hat{H},\hat{q}]=-i \frac{\hat{p}}{m}$, we then apply $\langle s|\cdots| r\rangle$ to both sides and obtain
\begin{equation}
{p_{s r}}=im E_{s r} q_{s r}.
\end{equation}
Substituting all the above results into the definitions in Eqs.~\eqref{eq:brs} and \eqref{e14}, we then get the coefficients
\begin{equation}
\label{e17}
\begin{split}
b_{r s}(t)=\delta_{r s} \cos (\omega t)\,,\qquad c_{r}(t)=\cos^2 (\omega t).
\end{split}
\end{equation}
Finally, the OTOC defined in \eqref{e22} for the normal harmonic oscillator is derived as
\begin{equation}\label{eq:CTHO}
C_{T}(t)=\cos ^{2}(\omega t).
\end{equation}
We note that the average of Eq.~\eqref{e21} is trivial because $c_{r}(t)$ is independent of $r$. Consequently, we can find that the OTOC for the harmonic oscillator, \ie $C_{T}(t)$ is extremely simplified since it is not involved with the temperature $T$.

We are more interested in studying the OTOC for the inverted harmonic oscillator whose Hamiltonian is shown in Eq.~\eqref{e1}, which is given by that of the harmonic oscillator via the analytical continuation $\omega \to i \omega$. By using the same trick, the OTOC for the inverted harmonic oscillator reads
\begin{equation}
\label{e4}
\begin{aligned}
C_{T}(t)=&-\left\langle[\hat{q}(t), \hat{p}(0)]^{2}\right\rangle_\beta= \cosh^2(\omega t) =
\frac{1}{2}+\frac{1}{4}e^{2\omega t}+ \frac{1}{4}e^{-2\omega t}\,,
\end{aligned}
\end{equation}
where the last term is suppressed with the time evolution. In comparison with the oscillating behavior of the OTOC for the normal harmonic oscillator, \ie \eqref{eq:CTHO}, the distinct feature of the OTOC for the inverted harmonic oscillator is its exponential increase with time, which is approximately described by \cite{Ali:2019zcj}
\begin{equation}\label{eq:CT}
C_{T}(t)\approx e^{2 \lambda_{L}\left(t-t_{*}\right)} +\frac{1}{2} + \mathcal{O}\left(e^{-2\omega t} \right) \,,
\end{equation}
where $\lambda_{L}=\omega$ denotes the Lyapunov exponent and $t_*=\frac{\log2}{\omega}$ is referred to as the scrambling time. Obviously, the Lyapunov exponent $\lambda_{L}$ for the inverted harmonic oscillator only depends on the frequency rather than on the temperature. Intuitively, the commutator $[\hat{q}(t), \hat{p}(0)]$ becomes the Poisson bracket $i \hbar\{q(t), p\}=i \hbar \frac{\partial q(t)}{\partial q(0)}$ in the semiclassical limit. This gives the dependence of the final position on small changes in the initial position, the classical diagnostic of the butterfly effect. Nearby trajectories in the chaotic system diverge exponentially, $\sim e^{\lambda_{L} t}$ where $\lambda_L$ is a Lyapunov exponent. For early times, the OTOC $C_T(t) \sim \alpha^2 e^{2 \lambda_{L} t}$ so $t_{*} \sim \frac{1}{\lambda_{L}} \log \frac{1}{\alpha}$.

However, we would like to remark that the physical meaning of the analytic continuation for the inverted harmonic oscillator is less clear because it makes the energy spectrum pure imaginary and the definition of the micro-canonical/thermal OTOC is thus ambiguous \cite{Hashimoto:2020xfr}. On the other hand, the unbounded bottom of the potential may result in the violation of the inequality that constrains the upper bound of the Lyapunov exponent, \ie
\begin{equation}
\label{e19}
\lambda_{\mathrm{OTOC}}(T) \leq c T, \quad
\end{equation}
where $c$ denotes the irrelevant numerical coefficient at the order $\mathcal{O}(1)$. In fact, this inequality is caused by the structure of the inverted harmonic oscillator potential and the quantum resolution condition to discriminate the local maximum by wave functions. The bound \eqref{e20} is for large $N$ theories while the inequality \eqref{e19} is for a single degree of freedom \cite{Hashimoto:2020xfr}. For the inverted harmonic oscillator with a higher frequency $\omega > c T$, the inequality in \eqref{e19} is explicitly violated. In the following sections, we study the chaotic behavior of the inverted harmonic oscillator by focusing on the complexity. We will show a similar conclusion as that from the OTOC given by the analytical continuation.

\subsection{Circuit Complexity and Inner Product of Gaussian States}
First, we briefly review the definition of circuit complexity. Readers who are interested in details can refer to \cite{Chapman:2018hou}. We start from a reference state $\left|\psi_{\mt{R}}\right\rangle$ and prepare a specific target state $\left|\psi_{\mt{T}}\right\rangle$ through the unitary evolution, namely
\begin{equation}
\left|\psi_{\mt{T}}\right\rangle=\hat{U}_{\mt{TR}}\left|\psi_{\mt{R}}\right\rangle.
\end{equation}
In order to define the circuit complexity from the chosen reference state to the target state, we need to fix the set of fundamental gates that can realize the unitary transformation $\hat{U}_{\mt{TR}}$. Correspondingly, a specific way to build the unitary operator $\hat{U}_{\mt{TR}}$ corresponds to the decomposition along a path integral on the group manifold, namely
\begin{equation}
\hat{U}_{\mt{TR}}=\overleftarrow{\mathcal{T}} \exp \left[i \int_{0}^{1} \mathrm{~d} t \, Y^I(t)\hat{M_I}\right],
\end{equation}
where $\overleftarrow{\mathcal{T}}$ indicates the path order, $\hat{M_I}$ denotes a set of generators dual to the fundamental gates, and $Y^I(t)$ represents the number of basic gates $\hat{M_I}$ used at ``time'' $t$ along the evolution path. While the circuit complexity still depends on the choice of the cost functions that parametrizes the cost of the various paths. The simple ones are given by the so-called $F_1$ and $F_2$ cost functions with the circuit complexity as
\begin{equation}
\label{e9}
\begin{split}
\mathcal{C}_1 &=\min _{Y^I(t)}\int_{0}^{1} \mathrm{~d} t \sum_I| Y^I(t)|\,,\\ \mathcal{C}_2&=\min _{Y^I(t)}\int_{0}^{1} \mathrm{~d} t \sqrt{\sum_{I}\left(Y^{I}(t)\right)^{2}}\,.\\
\end{split}
\end{equation}
According to the definition, the circuit complexity is the minimal cost of a given reference state to a chosen target state, \ie the cost of the optimal quantum circuit. Correspondingly, the minimization appearing in the expressions \eqref{e9} means the optimization of all possible paths connecting the reference state to the target state. In particular, the circuit complexity defined in \eqref{e9} associated with the $F_2$ cost function is nothing but the length of the shortest geodesic on the group manifold. However, it is not easy to precisely solve the optimization problem for a non-trivial group. The explicit realization of the calculations for the circuit complexity in the quantum field theory was first made for Gaussian states in \cite{Jefferson:2017sdb}. See \cite{Chapman:2018hou,Bhattacharyya:2019kvj,Hackl:2018ptj,Khan:2018rzm,Caceres:2019pgf,Bhattacharyya:2018bbv,Guo:2018kzl,Doroudiani:2019llj,Bai:2021ldj,Guo:2020dsi} for more following progress in this direction.

Due to the simplicity of Gaussian states $\ket{\psi}$ (see \eg \cite{Weedbrook:2015bva,wang2007quantum,adesso2012measuring,adesso2014continuous,ferraro2005gaussian}), one can show that they are completely characterized by the so-called covariance matrix, \ie the two-point functions defined by,
\begin{equation}
G^{a b} = \bra{\psi}(\hat{\xi}^{a} \hat{\xi}^{b}+\hat{\xi}^{b} \hat{\xi}^{a})\ket{\psi},
\end{equation}
where the vector operator $\hat{\xi}^{a}=\left(\hat{q} g,\frac{\hat{ p}}{g}\right)$. Note that $g$ is a new gate scale, and its dimension is one. We introduce $g$ to offset the dimension of $\hat{q}$ and $\hat{p}$ so that the dimension of $\hat{\xi}^{a}$ is zero. Considering the unitary transformation between Gaussian states, namely
\begin{equation}
\label{e24}
\begin{aligned}
\ket{\psi^{\prime}}&=\hat{U}\left|\psi\right\rangle
,\quad \hat{U}=e^{-\frac{i}{2} k_{a b}^{} \hat{\xi}^{a} \hat{\xi}^{b}},
\end{aligned}
\end{equation}
one can rewrite it as a unique transformation of their corresponding covariance matrices as follows \cite{Weedbrook:2015bva}
\begin{equation}
\label{Trans}
\begin{aligned}
G^{\prime a b} &=\bra{\psi^{\prime}}(\hat{\xi}^{a} \hat{\xi}^{b}+\hat{\xi}^{b} \hat{\xi}^{a})\ket{ \psi^{\prime}}=M_{\ c}^{a} G^{c d} M_{\ d}^{b},\end{aligned}
\end{equation}
where
\begin{equation}
\label{e25}
\begin{aligned}
M=e^K,\quad    K_{ \ b}^{a} =\Omega^{a c} k_{c b},  \quad \Omega^{a b}=\left(\begin{array}{cc}
0 & 1 \\
-1 & 0
\end{array}\right).
\end{aligned}
\end{equation}
As a consequence, we should expect that the circuit complexity of Gaussian states can be rewritten as a function of the covariance matrix. Particularly, it was found in \cite{Chapman:2018hou} that the information about the circuit complexity is encoded in the relative covariance matrix that is defined by
\begin{equation}
\Delta=G_{\mt{T}}G_{\mt{R}}^{-1}.
\end{equation}
where $G_{\mt{R}}$ and $G_{\mt{T}}$ denote the covariance matrix of the reference state and the target state, respectively. We note that the two eigenvalues of the relative covariance matrix $\Delta$ are reciprocal to each other and greater than zero. In the following, we denote the eigenvalue of greater than or equal to 1 as $\rho$. For Gaussian states, it was explicitly shown in Ref.~\cite{Chapman:2018hou} that the circuit complexity related to the $F_2$ cost function is given by
\begin{equation}
\mathcal{C}_{2}\left(G_{\mathrm{R}}, G_{\mathrm{T}}\right)=\frac{1}{2 \sqrt{2}} \sqrt{\operatorname{Tr}\left[(\log \Delta)^{2}\right]} =\frac{1}{2}\log{\rho}\,,
\end{equation}
In order to determine the equivalence between the covariance matrix method and the geometric method introduced before, we can take the one-mode Gaussian state as an example. In order to realize the unitary transformations between one-mode Gaussian states, we need to introduce three generators\footnote{In fact, $\hat{q},\hat{ p}$ here means that $G_R$ and $G_T$ are diagonal in $\hat{q},\hat{ p}$, but it does not represent the original $\hat{q},\hat{ p}$ \cite{Chapman:2018hou}.}, {\it viz},
\begin{equation}
\hat{M}_W=\frac{1}{2}(\hat{q}\hat{ p}+\hat{p}\hat{ q}), \quad \hat{M}_V=\frac{\hat{q}^{2}g^2}{\sqrt{2}}, \quad \hat{M}_Z=\frac{\hat{p}^{2}}{\sqrt{2}g^2},
\end{equation}
whose commutation relations precisely match the algebra $\mathfrak{s p}(2, \mathbb{R})$. It was shown in Ref.~\cite{Chapman:2018hou} that the shortest geodesic line for $\mathcal{C}_1$ and $\mathcal{C}_2$ is the straightline parametrized by
\begin{equation}
Y^{W}=-\frac{1}{2}\log\rho, \quad Y^{V}=0, \quad Y^{Z}=0.
\end{equation}
Substituting the solution of the optimal path into the definitions of the circuit complexity, \ie Eq.~\eqref{e9}, we easily obtain
\begin{equation}
\label{e28}
\begin{aligned}
\mathcal{C}_1&=
\int_{0}^{1} \mathrm{~d} t \sum_I| Y^I(t)|=\frac{1}{2}\log{\rho},   \\
\mathcal{C}_2 &=\int_{0}^{1} \mathrm{~d} t \sqrt{\sum_{I}\left(Y^{I}(t)\right)^{2}}=\frac{1}{2}\log{\rho},
\end{aligned}
\end{equation}
which is the same as that derived from the covariance matrix.

On the other hand, we are also interested in the inner product between the reference state and the target state. For the Gaussian states, the inner product also depends on the relative covariance matrix, \ie
\begin{equation}
\label{e8}
\mathcal{I}=\left|\left\langle G_{\mt{R}} \mid G_{\mt{T}}\right\rangle\right|^{2}=\operatorname{det} \frac{\sqrt{2} \Delta^{1 / 4}}{\sqrt{\mathbb{1}+\Delta}}=\frac{2 \sqrt{\rho}}{1+\rho}.
\end{equation}
From this viewpoint, we can find that the inner product of the Gaussian states encodes the same information as the circuit complexity between states.

\section{Chaos in Inverted Harmonic Oscillator}
In this section, we investigate the chaos in (inverted) harmonic oscillators by using the circuit complexity and the inner
product. Firstly, we choose the reference state as
\begin{equation}
\psi_{\mt{R}}\left(q\right)=\left(\frac{ a m\omega}{\pi}\right)^{1/4} \exp \left(- \frac{1}{2}a m\omega q^{2}\right),
\end{equation}
where $a$ is a dimensionless parameter that characterizes the Gaussian states. It is obvious that the reference state is nothing but the ground state of the harmonic oscillator with mass $am$. Correspondingly, its two-point function reads as
\begin{equation}
G_{\mt{R}}=\left(\begin{array}{cc}
\frac{g^2}{am\omega} &0\\
0 & \frac{am\omega}{g^2}  \\
\end{array}\right).
\end{equation}
We consider a particular target state $\psi_{\mt{T}}$ by a backward and forward time evolution from the reference state, \ie
\begin{equation}
\label{e30}
|\psi_{\mt{T}}\rangle=e^{i \hat{H}'  t}e^{-i \hat{H}  t} \left|\psi_{\mt{R}}\right\rangle\,.
\end{equation}
In particular, we consider $\hat{H}$ as the Hamiltonian of the system, and $\hat{H}'$ as the perturbed Hamiltonian with
\begin{equation}
\label{e26}
\hat{H}'=\hat{H}+\delta\hat{H}\,,\qquad  \delta\hat{H}\ll\hat{H}.
\end{equation}
Without the perturbation $\delta \hat{H}$, the target state trivially evolves back to the reference state after the backward and forward time evolutions. Therefore, the corresponding circuit complexity simply vanishes. By imposing the small perturbation $\delta \hat{H}$, one may expect that the circuit complexity is still small. However, we will show that the extremely small perturbation still could result in exponentially large circuit complexity for chaotic systems, \eg inverted harmonic oscillators. We will also show this chaotic behavior from the aspect of the inner product, which is also known as the Loschmidt echo.

\subsection{Harmonic oscillator}
\begin{figure}[ht!]
\centering
\includegraphics[width=3in]{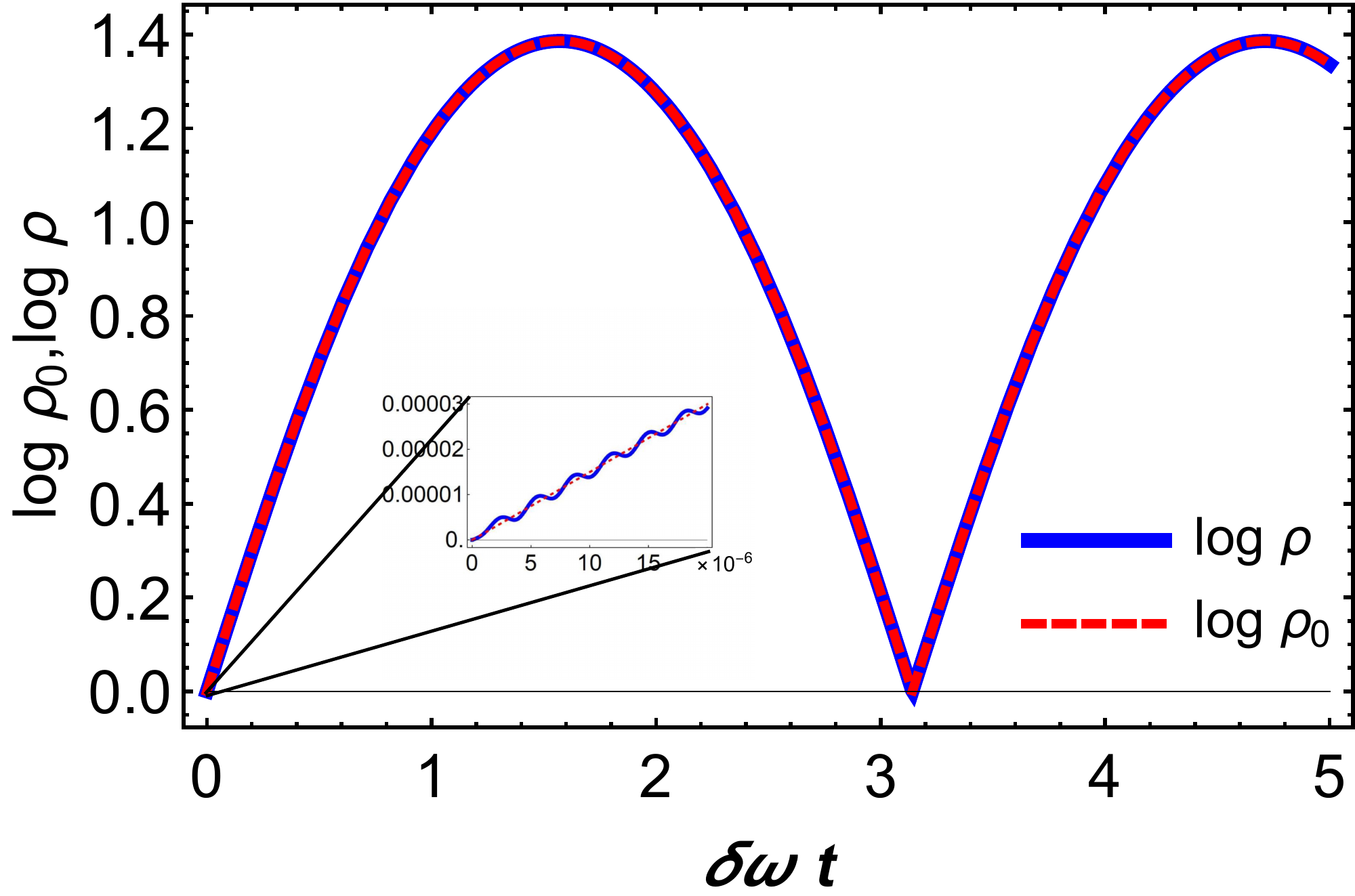}
\includegraphics[width=3in]{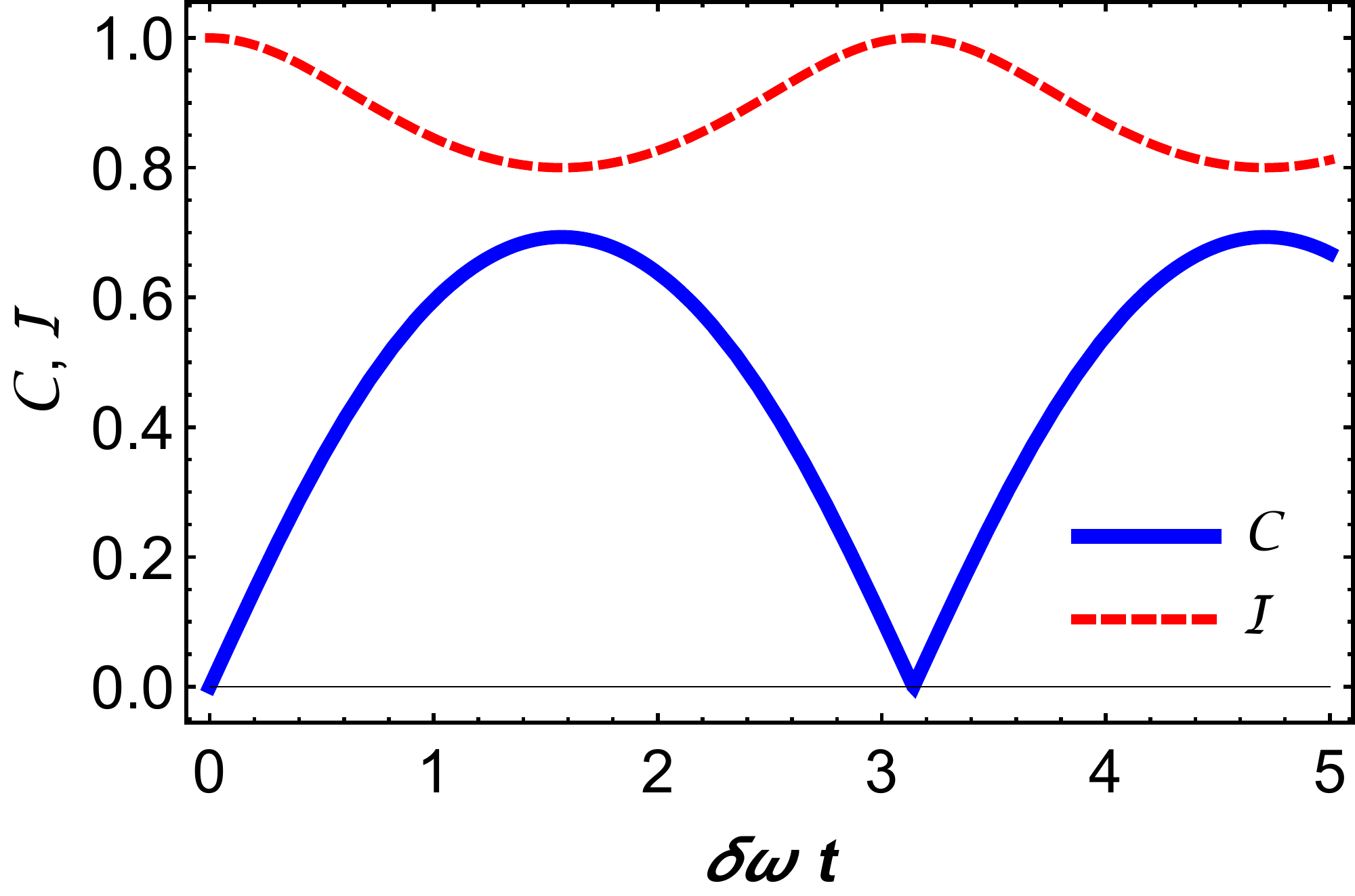}
\caption{Oscillating behavior for harmonic oscillators. Left: $\log\rho_0$ and $\log\rho$ vs time. Right: The circuit complexity $\mathcal{C}$ and the inner product $\mathcal{I}$ vs time. The parameters are set as $a=2$ and $\delta\omega/\omega=10^{-6}$. Note that the simple relation $\mathcal{C}=\mathcal{C}_1=\mathcal{C}_2$ for a single harmonic oscillator.}
\label{F6}
\end{figure}
In order to compare with the results of chaos, we begin with the simple case by considering the normal harmonic oscillator whose Hamiltonian is defined in Eq.~\eqref{eq:HOHamiltonian}. For the perturbed Hamiltonian, we focus on a simple case given by
\begin{equation}\label{eq:perturbedH}
\hat{H}'=\frac{\hat{p}^2}{2m}+\frac{1}{2} m (\omega+\delta\omega)^2 \hat{q}^2\quad  \text{with}\qquad  \frac{\delta\omega}{\omega} \ll 1\,.
\end{equation}
As introduced around Eq.~\eqref{e24}, we can parametrize the Gaussian unitary operators $e^{-i \hat{H}  t}$ and $e^{i \hat{H}' t}$ by the symmetric matrices $k_{ab}$, \ie
\begin{equation}
k=\left(\begin{array}{cc}
    \frac{\omega^2 m t}{g^2} &0\\
    0 & \frac{g^2 t}{m}  \\
    \end{array}\right),\quad
k'=\left(\begin{array}{cc}
    -\frac{(\omega+\delta\omega)^2 m t}{g^2} &0\\
    0 & -\frac{g^2 t}{m}  \\
\end{array}\right).
\end{equation}
respectively. Using the transformation of the covariance matrix shown in Eq.~\eqref{Trans}, we can obtain the covariance matrix of the target state \eqref{e30}, namely
\begin{equation}
G_{\mt{T}}=M'MG_{\mt{R}}M^T(M')^T,
\end{equation}
where
\begin{equation}
\begin{aligned}
M& =\left(\begin{array}{cl}
\frac{1}{2} e^{-i \omega t}\left(e^{2 i t \omega}+1\right) & -\frac{i e^{-i \omega t}\left(e^{2 i \omega t}-1\right) g^{2}}{2 m \omega} \\
\frac{i e^{-i \omega t}\left(e^{2 i \omega t}-1\right) m \omega}{2 g^{2}} & \frac{1}{2} e^{-i \omega t}\left(e^{2 i \omega t}+1\right)
\end{array}\right), \\
M'&=\left(\begin{array}{cc}
\frac{1}{2} e^{-i(\omega+\delta \omega)t}\left(e^{2 i(\omega+\delta \omega)t}+1\right) & \frac{i e^{-i(\omega+\delta \omega)t}\left(e^{2 i(\omega+\delta \omega)t}-1\right) g^{2}}{2 m(\omega+\delta \omega)} \\
-\frac{i e^{-i(\omega+\delta \omega)t}\left(e^{2 i(\omega+\delta \omega)t}-1\right) m(\omega+\delta \omega)}{2 g^{2}} & \frac{1}{2} e^{-i(\omega+\delta \omega)t}\left(e^{2 i(\omega+\delta \omega)t}+1\right)
\end{array}\right).
\end{aligned}
\end{equation}
For the sake of evaluating the circuit complexity of the target state \eqref{e30}, we recall that it is completely determined by the relative covariance matrix $\Delta\equiv G_{\mt{T}} G^{-1}_{\mt{R}}$.
By combining the expressions shown above, it is straightforward to derive the two eigenvalues of the relative covariance matrix, \ie $\rho, 1/\rho$. Taking the infinitesimal perturbation limit $\frac{\delta\omega}{\omega}\to0$, the leading term of the eigenvalue $\rho$ (denoted by $\rho_0$) is derived as\footnote{Note that $\rho$ does not depend on the gate scale $g$ and mass $m$.}
\begin{equation}\label{e27}
 \begin{split}
    \rho \approx \rho_0 =& \frac{(1+2 a^{2})^2-\left(a^{2}-1\right)^{2} \cos (2\delta\omega t )}{4 a^{2}} \\ &+\sqrt{\frac{\left(a^{2}-1\right)^{2}\left(1+6 a^2+a^4-\left(a^{2}-1\right)^{2} \cos (2\delta\omega t )\right) \sin^{2} (\delta\omega t )}{8 a^{4}}} \,.
 \end{split}
\end{equation}

Obviously, the above approximation implies that $\rho$ oscillates with time and the period is $\frac{\pi}{\delta\omega}$. Since $\delta\omega\ll\omega$, the period $\frac{\pi}{\delta\omega}$ is much larger than the oscillation time of the harmonic oscillator $\frac{2\pi}{\omega}$. In fact, $\rho$ is the superposition of two types of sine functions, namely $\frac{\delta\omega}{\omega}\sin (\omega t)$ and $\sin (\delta \omega t)$. The difference between $\log \rho, \log \rho_0$ is also presented in the left plot of Fig.~\ref{F6}. Substituting Eq.~\eqref{e27} into Eq.~\eqref{e28} and Eq.~\eqref{e8}, we can get the expression of complexity and inner product, respectively. Due to the oscillating behavior of $\rho$, one can expect that the circuit complexity and the inner product also have the same period $\frac{\pi}{\delta \omega}$. This result is also shown in the numerical plot in Fig.~\ref{F6}.


\subsection{Inverted harmonic oscillator}
Compared with the normal harmonic oscillator without any chaotic behaviors, the inverted harmonic oscillator presents distinguishable differences in terms of not only the circuit complexity but also the inner product. In order to explore that, we consider a similar perturbed Hamiltonian defined by
\begin{equation}\label{eq:perturbedH02}
\hat{H}'=\frac{\hat{p}^2}{2m}-\frac{1}{2} m (\omega+\delta\omega)^2 \hat{q}^2\,\quad  \text{with}\qquad  \frac{\delta\omega}{\omega} \ll 1\,,
\end{equation}
which can be also derived from Eq.~\eqref{eq:perturbedH} by the analytical continuation.
It is then straightforward to obtain the transformation matrices for the covariance matrix by
\begin{equation}
\begin{aligned}
M&=\left(\begin{array}{cc}
\frac{1}{2} e^{-\omega t}\left(e^{2 \omega t}+1\right) & \frac{e^{-\omega t}\left(e^{2 \omega t}-1\right) g^{2}}{2 m \omega} \\
\frac{e^{-\omega t}\left(e^{2 \omega t}-1\right) m \omega}{2 g^{2}} & \frac{1}{2} e^{-\omega t}\left(e^{2 \omega t}+1\right)
\end{array}\right),
\\
M'&=\left(\begin{array}{cc}
\frac{1}{2} e^{-t(\omega+\delta \omega)}\left(e^{2(\omega+\delta \omega)t}+1\right) & -\frac{e^{-t(\omega+\delta \omega)}\left(e^{2(\omega+\delta \omega)t}-1\right) g^{2}}{2 m(\omega+\delta \omega)} \\
-\frac{e^{-t(\omega+\delta \omega)}\left(e^{2(\omega+\delta \omega)t}-1\right) m(\omega+\delta \omega)}{2 g^{2}} & \frac{1}{2} e^{-t(\omega+\delta \omega)}\left(e^{2(\omega+\delta \omega)t}+1\right)
\end{array}\right),
\end{aligned}
\end{equation}
with respect to the unitary transformation operations from the backward and forward time evolutions. As a result, one can further obtain the covariance matrix of the target state $G_{\mt{T}}=M'MG_{\mt{R}}M^T(M')^T$ and the corresponding relative covariance matrix $\Delta=G_{\mt{T}}G_{\mt{R}}^{-1}$ with complicated but analytic expressions. Of course, all information has been encoded in the eigenvalue of $\Delta$, \ie $\rho$ which is greater than or equal to 1.

\begin{figure}[ht!]
\centering
\includegraphics[width=8cm]{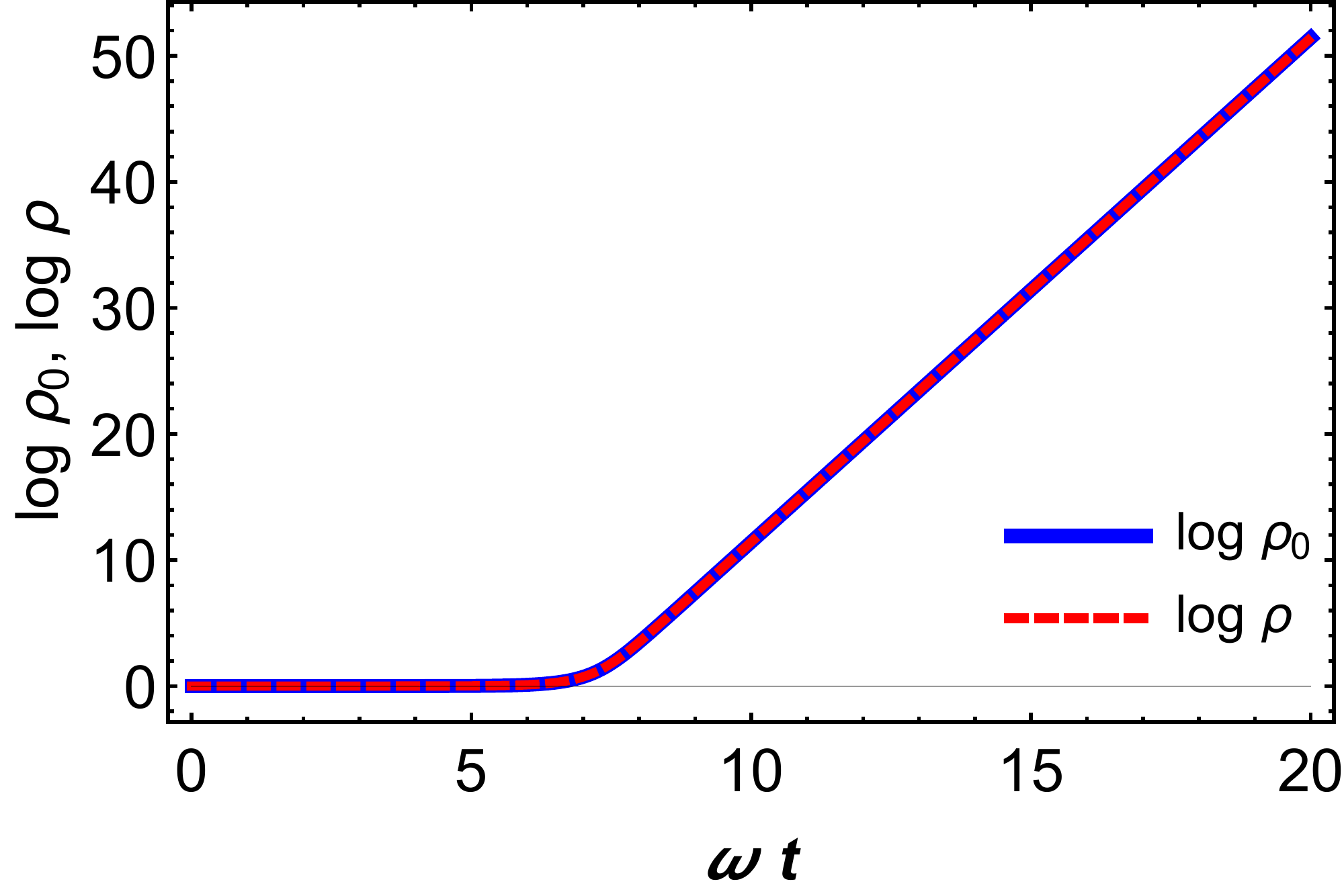}
\caption{$\log\rho_0$ and $\log\rho$ vs time. The parameters are set as $a=2$ and $\delta\omega/\omega=10^{-6}$. }
\label{F1}
\end{figure}

In order to obtain a simple analytical expression of the chaotic behaviors, we focus on the infinitesimal perturbation by taking $\frac{\delta\omega}{\omega}\to0$ to simplify the precise expression of $\rho$. Finally, we find that the leading terms of $\rho$ are derived as
\begin{equation}
\label{e29}
\begin{aligned}
\rho \approx \rho_0=&  1+ \frac{\delta \omega^2}{\omega^2} \left( \frac{\left(1+a^{2}\right)^{2}  \cosh (4 \omega t)}{16 a^{2} }-\cosh (2 \omega t)\right)
\\& +\frac{\delta \omega}{\omega} \left(\frac{\left(1+a^{2}\right)^{2}  \cosh (4 \omega t)}{8 a^{2} }-2\cosh (2 \omega t) \right.
\\&  \left. -\frac{\left(1+a^{2}\right)^{2} \delta \omega^{2} \cosh (6 \omega t)}{16 a^{2} \omega^{2}}+\frac{\left(1+a^{2}\right)^{4} \delta \omega^{2} \cosh (8 \omega t)}{512 a^{4} \omega^{2}}\right)^{1 / 2}\,,
\end{aligned}
\end{equation}
where we still keep higher-order terms $\delta \omega^2, \delta \omega^4$ due to the competing factors $\cosh (\#\omega t)$ with exponentially increases in time\footnote{Note that $\rho_0$ is a complex number in some cells, this is because we ignore the high-order small quantities under the radical sign.}. It is worth mentioning that through analytic continuation ($a \to i a,\omega \to i \omega,\delta\omega \to i \delta\omega$), $\rho$ in Eq.~\eqref{e27} and $\rho$ in Eq.~\eqref{e29} are the reciprocal of each other. This is because the inverted harmonic oscillator can be regarded as the analytical extension of the harmonic oscillator. Comparison between $\log \rho_0$ and the precise value of $\log\rho$ is shown in Fig.~\ref{F1}.

As expected, the periodic behavior appearing on the normal harmonic oscillator is replaced by the exponential increase for the inverted harmonic oscillator. Note that there are four modes of exponential change over time, namely $\omega,2\omega,3\omega,4\omega$. At early times, the exponential increase is dominated by
\begin{equation}
\begin{aligned}
\rho\approx &\frac{1+\cosh(2\delta\omega t )}{2}+\frac{a^2\left(\cosh(2\delta\omega t )-1\right)}{4}+\frac{\cosh(2\delta\omega t )-1}{4a^2}
\\&+\frac{1-e^{-2\delta\omega t }}{8 a^{2}}  \left(1+a^{2}\right) \sqrt{\left(1+a^{2}\right)^{2}-2\left(1-6 a^{2}+a^{4}\right) e^{2\delta\omega t }+\left(1+a^{2}\right)^{2} e^{4\delta\omega t }}.
\end{aligned}
\end{equation}
Unlike Eq.~\eqref{e29} and similar to Eq.~\eqref{e27}, only the zero-order term of $\frac{\delta\omega}{\omega}$ is retained in the above formula. In this range, $\rho$ is approximately one and consistent with the early behavior in Fig.~\ref{F1}. At late times, the exponential increase is dominated by
\begin{equation}
\rho\approx \frac{\left(1+a^{2}\right)^{2} e^{4 \omega t} \delta \omega^{2}}{16 a^{2} \omega^{2}}  .
\end{equation}

First of all, we can apply the formula \eqref{e8} to the Gaussian states to evaluate the inner product between our reference state and the target state at late times, \ie
\begin{equation}
\label{e11}
-\log\left|\left\langle G_{\mathrm{R}} \mid G_{\mathrm{T}}\right\rangle\right|^{2}=-\log\frac{2 \sqrt{\rho}}{1+\rho}\approx {2\omega  \left(t-\frac{1}{2\omega}\log\frac{8a\omega}{(1+a^2)\delta\omega}\right)} \,,
\end{equation}
whose linear growth characterizes the chaos of the inverted harmonic under perturbations. The maximal Lyapunov exponent is expected to be related to the Loschmidt echo $\mathcal{I}(t)$ in terms of  \cite{Jalabert2001}
\begin{equation}
\lambda^{\mathcal{I}}_L=-\lim_{t\rightarrow \infty}\frac{1}{t}\log\mathcal{I}(t).
\end{equation}
For the inverted harmonic oscillator, we simply have $\lambda^{\mathcal{I}}_L= 2\omega$ by using the late-time approximation \eqref{e11}. This is obviously inconsistent with the result derived from the OTOC in Eq.~\eqref{eq:CT}, \ie $\lambda^{\mathcal{I}}_L=2\lambda_L$. This shows that OTOC and Loschmidt echo describe different aspects of quantum chaos.

On the other hand, we are also interested in circuit complexity, which is defined in \eqref{e28} for a single Gaussian mode. Taking the late-time limit, we simply get the circuit complexity of the inverted harmonic oscillator, namely
\begin{equation}
\begin{aligned}
\label{e5}
\mathcal{C}(t)=\mathcal{C}_1(t)=\mathcal{C}_2(t)=\frac{1}{2}\log\rho\approx 2\omega\left(t -  \frac{1}{2\omega}\log \frac{4 a \omega}{(1+a^2) \delta \omega} \right) \,,
\end{aligned}
\end{equation}
which presents a similar linear growth\footnote{Note that the complexity here increases linearly with time and hence has no upper bound. However, the complexity calculated by different methods in Ref.~\cite{Bhattacharyya:2020iic} has an upper bound.} as $-\log\mathcal{I}(t)$. Correspondingly, we find that the circuit complexity also probes the chaos of the inverted harmonic. For example, we can find that the growth rate of the circuit complexity is proportional to the Lyapunov exponent of the classical inverted oscillator, \ie
\begin{equation}
\lambda^{\mathcal{C}}_L=\lim_{t\rightarrow \infty}  \frac{\mathcal{C}(t)}{t}  = 2\omega =2 \lambda_{L}.
\end{equation}
The same result was also derived in Ref.~\cite{Ali:2019zcj} by fitting the numerical results.

Except for the growth rate, another characteristic quantity in the expression of the circuit complexity is the time delay, \ie
\begin{equation}
t_s^{\mathcal{C}}= \frac{1}{2\omega}\log \frac{4 a \omega}{(1+a^2) \delta \omega} = \frac{t_*}{2}+\frac{1}{2\omega}\log\frac{2a\omega}{(1+a^2)\delta\omega}\,.
\end{equation}
It represents the time scale when the circuit complexity starts increasing linearly, but it is obviously different from the scrambling time $t_\ast$.
It is worth pointing out that $t^{\mathcal{C}}_s$ increases with the decrease of the amplitude of the perturbation, \ie $\delta \omega$. This behavior is expected because the system takes a longer time to respond to a smaller perturbation.

Similarly, we can find that the Loschmidt echo $ \mathcal{I}(t)$ shows a similar response time, namely
\begin{equation}
\label{e31}
t^{\mathcal{I}}_s=\frac{1}{2\omega}\log\frac{8a\omega}{(1+a^2)\delta\omega}.
\end{equation}
It can be seen that the rate of the inner product decrease is consistent with the rate of the complexity increase, but the time when the inner product starts to decrease is later than the time when the complexity increases. We finally point out that the rate of change of the complexity and inner product does not depend on the choice of the reference state, but the time to start the change depends on the choice of the reference state. That is, $t^{\mathcal{C}}_s$ an $t^{\mathcal{I}}_s$ both depend on the dimensionless scale $a$ of the reference state and both reach the maximum at $a=1$ where the reference state is the ground state of the harmonic oscillator with the same frequency $\omega$.

\section{Chaos in Quantum Field Theory}
In this section, we investigate the chaos in quantum field theory by using circuit complexity and Loschmidt echo. In particular, we consider the theory of two free scalar fields ((1+1) -dimensional field theory) with the following Hamiltonian \cite{Ali:2019zcj}
\begin{equation}
 \hat{H}=\hat{H}_{0}+\hat{H}_{I},
\end{equation}
where
\begin{equation}
\begin{aligned}
\hat{H}_{0} &=\frac{1}{2} \int d x\left[\hat{\Pi}_{1}^{2}+\left(\partial_{x} \hat{\phi}_{1}\right)^{2}+\hat{\Pi}_{2}^{2}+\left(\partial_{x} \hat{\phi}_{2}\right)^{2}+m^{2}\left(\hat{\phi}_{1}^{2}+\hat{\phi}_{2}^{2}\right)\right],\\
 \hat{H}_{I} &= \frac{\lambda}{2}
        \int d x \partial_{x} \hat{\phi}_{1}
                 \partial_{x} \hat{\phi}_{2}.
\end{aligned}
\end{equation}
It can be discrete by placing it on the lattice, \ie
\begin{align}
\hat{H}=&\frac{\delta}{2}\sum_{n=0}^{N-1}\left[\hat{p}_{1, n}^{2}+\hat{p}_{2, n}^{2}+\Omega^{2}\left(\hat{x}_{1, n+1}-\hat{x}_{1,n}\right)^{2}+\Omega^{2}\left(\hat{x}_{2, n+1}-\hat{x}_{2, n}\right)^{2}\right. \\
&+\left.\tilde{m}^{2}\left(\hat{x}_{1, n}^{2}+\hat{x}_{2, n}^{2}\right)+\tilde{\lambda}\left(\hat{x}_{1, n+1}-\hat{x}_{1, n}\right)\left(\hat{x}_{2, n+1}-\hat{x}_{2, n}\right)\right],\label{e13}
\end{align}
with
\begin{equation}
\hat{x}(\vec{n})=\delta \hat{\phi}(\vec{n}),\quad
\hat{p}(\vec{n})=\frac{\hat{\Pi}(\vec{n})}{\delta},\quad
\Omega=\frac{1}{\delta^{2}},\quad
\tilde{\lambda}=\frac{\lambda}{ \delta^{4}},\quad
\tilde{m}=\frac{m}{\delta},
\end{equation}
where $\delta$ is the lattice spacing and $N$ is the number of lattice points. In order to release the coupling of different degrees of freedom, we perform the following transformations
\begin{equation}
\label{e15}
\begin{array}{l}
\hat{x}_{1, l}=\frac{1}{\sqrt{N}} \sum_{k=0}^{N-1} \exp \left(\frac{2 \pi i k}{N} l\right) \hat{x}'_{1, k}, \\
\hat{p}_{1, l}=\frac{1}{\sqrt{N}} \sum_{k=0}^{N-1} \exp \left(-\frac{2 \pi i k}{N} l\right) \hat{p}'_{1, k}, \\
\hat{x}_{2, l}=\frac{1}{\sqrt{N}} \sum_{k=0}^{N-1} \exp \left(\frac{2 \pi i k}{N} l\right) \hat{x}'_{2, k}, \\
\hat{p}_{2, l}=\frac{1}{\sqrt{N}} \sum_{k=0}^{N-1} \exp \left(-\frac{2 \pi i k}{N} l\right) \hat{p}'_{2, k}, \\
\end{array}
\end{equation}
and
\begin{equation}
\begin{array}{l}
\hat{p}'_{1, k}=\frac{\hat{p}_{s, k}+\hat{p}_{a, k}}{\sqrt{2}},\quad \hat{p}'_{2, k}=\frac{\hat{p}_{s, k}-\hat{p}_{a, k}}{\sqrt{2}}, \\
\hat{x}'_{1, k}=\frac{\hat{x}_{s, k}+\hat{x}_{a, k}}{\sqrt{2}},\quad \hat{x}'_{2, k}=\frac{\hat{x}_{s, k}-\hat{p}_{a, k}}{\sqrt{2}},
\end{array}
\end{equation}
which lead to the Hamiltonian
\begin{equation}
\label{e16}
\hat{H}= \sum_{k=0}^{N-1} \left[\frac{\hat{p}_{s, k}^{2}}{2m_k}+\frac{1}{2}m_k\bar{\omega}_{k}^{2} \hat{x}_{s, k}^{2}+\frac{\hat{p}_{a, k}^{2}}{2m_k}+\frac{1}{2}m_k\omega_{k}^{2} \hat{x}_{a, k}^{2}\right],
\end{equation}
where
\begin{eqnarray}
m_k &=&\frac{1}{\delta}, \nonumber \\
\omega_k&=&\sqrt{\frac{4(1-\lambda)}{\delta^2}
\sin^2\left(\frac{\pi k}{N}\right)+m^2},  \\
\bar{\omega}_k&=&\sqrt{\frac{4(1+\lambda)}{\delta^2}
\sin^2\left(\frac{\pi k}{N}\right)+m^2}. \nonumber
\end{eqnarray}
The position and momentum operators obey the commutation relation
\begin{equation}
\left[\hat{x}_{s,k}, \hat{p}_{s,l}\right]=i\delta_{kl},\quad \left[\hat{x}_{a,k}, \hat{p}_{a,l}\right]=i\delta_{kl}.
\end{equation}
For simplicity, we take $m=0,\,\lambda>1$ in the following part. Obviously, in this case, $\bar{\omega}_{k}^{2}$ is always positive and $\omega_{k}^{2}$ is always negative for $k>0$. This means that we can divide the system into three parts: harmonic oscillators, inverted harmonic oscillators, and free particles (for $k=0$). Since we are interested in the unstable behavior of the system, we mainly examine the inverted oscillator part with the Hamiltonian
\begin{equation}
\hat{H}'= \sum_{k=1}^{N-1} \left[\frac{\hat{p}_{a, k}^{2}}{2m_k}+\frac{1}{2}m_k\omega_{k}^{2} \hat{x}_{a, k}^{2}\right].
\end{equation}
Similar to the previous section, we first select the reference state as
\begin{equation}
\psi_{R}\left(x\right)=\otimes_k\left(\frac{m_k\vert\omega_k\vert}{\pi}\right)^{1/4} \exp {\left(- \frac{1}{2} m_k\vert\omega_k\vert x_{a,k}^{2}\right)},
\end{equation}
which is the ground state of the harmonic oscillator with mass $m_k$ and frequency $\vert\omega_k\vert$. Then we use the Hamiltonian $\hat{H}'(\lambda)$ to evolve forward and the perturbed Hamiltonian $\hat{H}'(\lambda+\delta\lambda)$ to evolve backward to get the target state. Finally, we get the complexity and the Loschmidt echo at late times, \ie
\begin{equation}
\begin{aligned}
\mathcal{C}_1&=\sum_{k=1}^{N-1}(\mathcal{C}_1)_k=\sum_{k=1}^{N-1}\left(2\omega_kt - \log \frac{2}{\alpha} \right)
\\&=\frac{4\sqrt{\lambda-1}}{\delta}\cot\left(\frac{\pi}{2N}\right)\left(t-\frac{\delta \tan\left(\frac{\pi}{2N}\right)\log\left(\frac{2}{\alpha}\right)(N-1)}{4\sqrt{\lambda-1}}\right),
\\
-\log\mathcal{I}&=
-\sum_{k=1}^{N-1}\log\mathcal{I}_k=\sum_{k=1}^{N-1}\left(2\omega_kt - \log \frac{4}{\alpha} \right)\\&
=\frac{4\sqrt{\lambda-1}}{\delta}\cot\left(\frac{\pi}{2N}\right)\left(t-\frac{\delta \tan\left(\frac{\pi}{2N}\right)\log\left(\frac{4}{\alpha}\right)(N-1)}{4\sqrt{\lambda-1}}\right),
\end{aligned}\end{equation}
where
\begin{equation}
\alpha=\frac{\delta\omega_k}{\omega_k}=
\frac{\frac{2\sqrt{\lambda+\delta\lambda-1}\sin(\frac{\pi k}{N})}{\delta}-\frac{2\sqrt{\lambda-1}\sin(\frac{\pi k}{N})}{\delta}}{\frac{2\sqrt{\lambda-1}\sin(\frac{\pi k}{N})}{\delta}}=\sqrt{\frac{\lambda+\delta\lambda-1}{\lambda-1}}-1.
\end{equation}
As in the previous section, we can describe the chaotic behavior with complexity and Loschmidt echo, \ie
\begin{equation}
\begin{aligned}
\lambda^{\mathcal{C}}_L&=\frac{4\sqrt{\lambda-1}}{\delta}\cot\left(\frac{\pi}{2N}\right),\quad t^{\mathcal{C}}_s=\frac{\delta \tan\left(\frac{\pi}{2N}\right)\log\left(\frac{2}{\alpha}\right)(N-1)}{4\sqrt{\lambda-1}},\\
\lambda^{\mathcal{I}}_L&=\frac{4\sqrt{\lambda-1}}{\delta}\cot\left(\frac{\pi}{2N}\right),\quad t^{\mathcal{I}}_s=\frac{\delta \tan\left(\frac{\pi}{2N}\right)\log\left(\frac{4}{\alpha}\right)(N-1)}{4\sqrt{\lambda-1}}.
\end{aligned}
\end{equation}
Due to quantum field theory is the continuity limit of the lattice theory, below we give the values of the above physical quantities under the continuity limit, \ie
\begin{itemize}
  \item when $N\to+\infty$, 
  \begin{equation}
     \lambda^{\mathcal{C}}_L\to+\infty, \quad
     t^{\mathcal{C}}_s\to
        \frac{\pi\delta \log\left(\frac{2}{\alpha}\right)}
             {8\sqrt{\lambda-1}},\quad
     \lambda^{\mathcal{I}}_L\to+\infty, \quad
     t^{\mathcal{I}}_s\to
        \frac{\pi\delta \log\left(\frac{4}{\alpha}\right)}
             {8\sqrt{\lambda-1}};
  \end{equation}
  \item when $\delta\to0$, 
  \begin{equation}
       \lambda^{\mathcal{C}}_L\to+\infty, \quad
       t^{\mathcal{C}}_s\to0,\quad
       \lambda^{\mathcal{I}}_L\to+\infty, \quad
       t^{\mathcal{I}}_s\to0;
  \end{equation}
  \item when $\delta\to0$  and $N\to+\infty$, 
  \begin{equation}
        \lambda^{\mathcal{C}}_L\to+\infty, \quad
        t^{\mathcal{C}}_s\to 0,\quad
        \lambda^{\mathcal{I}}_L\to +\infty, \quad
        t^{\mathcal{I}}_s\to 0.
  \end{equation}
\end{itemize}


\section{Summary}
In this article, we investigated the chaotic behaviors of the inverted harmonic oscillator through the circuit complexity and Loschmidt echo. Using the covariance matrix associated with Gaussian states, we obtained the analytical expressions of the circuit complexity and Loschmidt echo. We found that the leading contributions of the circuit complexity $\mathcal{C}$ and Loschmidt echo $\mathcal{I}$ exhibit similar characteristics at late times,
\begin{equation}
\begin{aligned}
\mathcal{C}(t)&\approx  2\omega\left(t -  \frac{1}{2\omega}\log \frac{4 a \omega}{(1+a^2) \delta \omega} \right),  \\ \mathcal{I}(t)&\approx 2\omega\left(t -  \frac{1}{2\omega}\log \frac{8 a \omega}{(1+a^2) \delta \omega} \right)\,.
\end{aligned}\end{equation}
Through the above analytical results, we can read the Lyapunov exponents $\lambda^{\mathcal{C}}_L$ and $\lambda^{\mathcal{I}}_L$ and the scrambling times $t^{\mathcal{C}}_s$ and $t^{\mathcal{I}}_s$. The Lyapunov exponent $\lambda^{\mathcal{C}}_L$ is consistent with the result in \cite{Ali:2019zcj} that was derived by using numerical fitting. Furthermore, we found that the scrambling time will diverge as the perturbation approaches zero because it takes infinite time for the system to respond to infinitesimal perturbations. We also found that the scrambling time explicitly depends on the choice of the reference states. It is obviously different from the Lyapunov exponent which is determined by the system itself and irrelevant to the choice of the reference states.

In the context of AdS/CFT duality, there were several proposals for the holographic dual of the circuit complexity, see \eg \cite{Susskind:2014moa,Susskind:2014rva,Stanford:2014jda,Brown:2015lvg,Brown:2015bva}. It would be interesting to investigate the chaotic behaviors of complexity from the viewpoint of holographic complexity. In particular, it is worth pointing out that the Lyapunov exponent of a quantum chaotic system is expected to be bounded above by $\frac{2\pi}{\beta}$, which is saturated by strongly-coupled field theories with a holographic dual. However, this contradicts the result for the inverted harmonic oscillator whose Lyapunov exponent linearly depends on its frequency.

\section*{Acknowledgments}
We would like to thank Shan-Ming Ruan for his very useful suggestions, comments, and help. We also thank Bin Guo, Jun-Jie Wan, and Jing-Cheng Chang for helpful discussions. This work was supported in part by the National Natural Science Foundation of China (Grants no. 11875151) and the 111 Project (Grant no. B20063).

\end{document}